\documentclass[onecolumn,groupedaddress,superscriptaddress,showpacs,showkeys,preprint]{revtex4-1}  

\usepackage[pdftex]{graphicx}                                		
\usepackage{amssymb,amsfonts,amsmath}                        		
\usepackage{textcomp,textgreek}				     		
\usepackage{bm}                                              		
\usepackage{dcolumn}                                         		
\usepackage{color,colordvi}                                  		
\usepackage{bbm}                                             		

\newcommand{\tdow}[2]{{#1}_{\mbox{\tiny $#2$}}}		     		
\newcommand{\tupp}[2]{{#1}^{\mbox{\tiny $#2$}}}		     		
\newcommand{\tsubb}[3]{{#1}_{\mbox{\tiny $#2$}}^{\mbox{\tiny $#3$}}}	
\newcommand{\Dbra}[1]{\left\langle#1\right|}                 		
\newcommand{\Dket}[1]{\left|#1\right\rangle}                 		
\newcommand{\Qcommu}[2]{[#1,#2]}                             		
\newcommand{\Acommu}[2]{\{#1,#2\}}                           		
\newcommand\id{\ensuremath{\mathbbm{1}}}         			

\begin{document}


\title{Biexponential decay and ultralong coherence of a single qubit}

\author{J\'{e}r\^{o}me Flakowski}
\affiliation{Polymer Physics, ETH Z\"{u}rich, Department of Materials, CH-8093 Z\"{u}rich, Switzerland}

\author{Maksym Osmanov}
\affiliation{Polymer Physics, ETH Z\"{u}rich, Department of Materials, CH-8093 Z\"{u}rich, Switzerland}

\author{David Taj}
\affiliation{Institute for Theoretical Physics, ETH Z\"{u}rich, Department of Physics, CH-8093 Z\"{u}rich, Switzerland}

\author{Hans Christian \"{O}ttinger}
\email{Electronic address: hco@mat.ethz.ch}
\affiliation{Polymer Physics, ETH Z\"{u}rich, Department of Materials, CH-8093 Z\"{u}rich, Switzerland}


\begin{abstract}
A quantum two-state system, weakly coupled to a heat bath, is traditionally studied in the Born-Markov regime under the secular approximation with completely positive linear master equations. Despite its success, this microscopic approach exclusively predicts exponential decays and Lorentzian susceptibility profiles, in disagreement with a number of experimental findings. To leave this limited paradigm, we use a phenomenological positive nonlinear master equation being both thermodynamically and statistically consistent. We find that, beyond a temperature-dependent threshold, a bifurcation in the decoherence time $T_2$ takes place; it gives rise to a biexponential decay and a susceptibility profile being neither Gaussian nor Lorentzian. This implies that, for suitable initial states, a major prolongation of the coherence can be obtained in agreement with recent experiments. Moreover, $T_2$ is no longer limited by the energy relaxation time $T_1$ offering novel perspectives to elaborate devices for quantum information processing.
\end{abstract}






\maketitle


\section{Introduction}
Duration of energy relaxation and decoherence is of significance for a wide scope of quantum nanodevices. Preserving their coherence is a particularly challenging task in the presence of noisy environments~\cite{Ladd2010}. The archetypical example is a qubit whose coherence time must be longer than the duration of a logic-gate operation to adequately carry out a quantum computation~\cite{Nielsen1997}. It also plays a central role in long distance quantum communication~\cite{Duan2001}, environment-assisted transport~\cite{Rebenrost2009,Marais2013}, long-lived coherence of photosynthetic complexes~\cite{Lee2007,Scholes2011}, quantum chaos~\cite{Pattanayak1997,Wu2009} and others~\cite{Mukamel1995,Breuer2002,Alicki2007,Weiss2008}. Theoretically, the energy relaxation and decoherence lifetimes, respectively denoted by $T_1$ and $T_2$, are often computed in the context of open quantum systems from the celebrated Lindblad-Davies master equation~(LDME)~\cite{Davies1974,Lindblad1976,Gorini1976}. This equation is obtained in the \textit{weak-coupling limit}~(WCL), where the coupling constant of the system-bath interaction is taken towards zero after a time rescaling. Although the LDME is related to an underlying Hamiltonian description only in this scaling limit, its linear and robust thermodynamic character~\cite{Spohn2007} makes it an appealing tool to compute lifetimes at small but finite values of the coupling constant. However, due to its structural properties the LDME exclusively predicts polarization decays of exponential kind, corresponding to Lorentzian profiles for the susceptibility. Precisely such features are violated in a plethora of experimental findings ranging from electron/nuclear spins~\cite{Sasaki2005,Shankar2010,Jarvinen2014,Gumann2014} or nitrogen-vacancy~(NV) centers~\cite{Beha2012,Grezes2014} to chromophoric molecules~\cite{Muller2010,SchlauCohen2012} which display a biexponential decay of the polarizations leading to two $T_2$ times, a short and a long one. This decay process was clearly associated to homogeneous non-Lorentzian susceptibility profiles in quantum dots~\cite{Birkedal2001,Borri2001,Borri2003,Vagov2004,Borri2007,Robert2012,Birindelli2014} and NV centers~\cite{Balasubramanian2009,Smith2011}. In addition, using materials doped with rare-earth ions, the decoherence can be slowed down by one order of magnitude for an initial Bloch vector being properly sized and oriented~\cite{MacFarlane1980,Fraval2005,McAuslan2012,PascualWinter2012}. On the other hand, the complete positivity of the LDME implies that $T_2$ is at most twice as large as $T_1$, while equality is reached only in the absence of pure dephasing~\cite{Gorini1978,Pottinger1985}. Although no experimental evidence has yet broken the theoretical bound $T_2 \leq 2T_1$ it has been highly disputed, see \cite{Laird1991a,Laird1991b,Laird1994,Budimir1987,Sevian1989,Aihara1990,Chang1993,Andreozzi1992,Reichman1996,Wonderen2000,Goun2002}. Indeed, the LDME is just one possible phenomenological Markovian linear master equation for finite couplings whose complete positivity is by far too restrictive~\cite{Majewski1990,Alicki1990,Pechukas1994,Pechukas1995,Czachor1998,Kimura2002,Shaji2005,McCracken2014}. To go beyond this strict picture one can resort to stochastic models~\cite{Sillescu1971,Donati1996,Kimura2002_strong}, non-Markovianity~\cite{Daffer2004,DiVincenzo2005,Krimer2014} or nonlinear dynamics~\cite{Willis1974,Grabert1974,Grabert1976,Grabert1977,Grabert1982,Becker1986,Bonci1991,Grigolini1991,Linden1998,Wonderen2000,Romero2004,Tsekov2009,Ottinger2010_TLS_DHO,Ottinger2011,Kominis2011} for the reduced~(open) quantum system.

Hereafter, we describe the emergence of two $T_2$ decoherence times, namely a short and a long one, for a qubit undergoing a physically sound Markovian dynamics beyond the WCL~\footnote{The coupling strength is non-vanishing, contrary to the WCL, and is ranging from the small to moderate/intermediate values~\cite{Breuer2002}.}. Thereby, the polarizations follow a biexponential decay coming along with a non-Lorentzian susceptibility profile. In this context, we explain how an appropriate choice of the initial state can slow down the decay of the polarizations which in turn can be exploited to overcome the well-known $T_2 \leq 2T_1$ bound.

\section{Traditional paradigm} 
Let us consider a qubit connected to a heat bath at inverse temperature $\beta$ described by a $2\times2$ density matrix $\rho$~(with the natural units $k_{\rm B}\!=\!1$ and $\hbar\!=\!1$). In a conventional approach, the dynamics of the system is modelled with the LDME~\cite{Davies1974,Breuer2002}
\begin{equation}
  \dot{\rho} = -i\Qcommu{\tdow{H}{\rm S}}{\rho}
  + \sum_\omega h(\omega) \left( A_\omega \rho A_\omega^\dagger - \frac{1}{2} \Acommu{A_\omega^\dagger A_\omega}{\rho} \right).
  \label{LDME}
\end{equation}
The first contribution, associated to the system Hamiltonian $\tdow{H}{\rm S}$, produces a reversible time-evolution whereas the second one, expressed in terms of the so-called Lindblad eigenoperators $A_\omega$ and a spectral function $\mbox{\small $h(\omega)$}$, induces relaxation and decoherence processes. The connection to an underlying Hamiltonian dynamics is established by linking the $A_\omega$ and $\mbox{\small $h(\omega)$}$ to the total Hamiltonian $\tdow{H}{}\!=\!\tdow{H}{\rm S}\otimes\tdow{\id}{\rm B} + \tdow{\id}{\rm S}\otimes \tdow{H}{\rm B} + Q\otimes\Phi$, where $\tdow{H}{\rm B}$ is the bath Hamiltonian while $Q$ and $\Phi$ are, respectively, the system and bath self-adjoint coupling operators. More concretely, the $A_\omega$ satisfy $[A_\omega,\!\tdow{H}{\rm S}] \!=\! \omega A_\omega$ and are provided in terms of $Q$ through the Kronecker delta relations $(A_\omega)_{ij} \!=\! Q_{ij} \, \delta_{\rm Kr}{(E_j-E_i,\omega)}$ using the matrix elements' notation $(\cdot)_{ij}\!=\! \Dbra{E_i} \!\cdot\! \Dket{E_j}$ in the system energy eigenbasis $\{\Dket{E_i}\}$. On the other hand, the spectral function $\mbox{\small $h(\omega)$} \!=\! \int_0^{\infty}\! dt \,  e^{i\omega{t}} \, {\rm tr}( \Phi(t) \Phi(0) \tdow{\pi}{B} )\!\geq{\!0}$, where $\tdow{\pi}{B}$ is the bath equilibrium state, respects the KMS condition $\mbox{\small $h(\omega)$} \!=\! e^{\beta\omega}\mbox{\small $h(-\omega)$}$~\cite{Alicki2007} implying that the system converges towards the Gibbs state $\pi = e^{-\beta\tdow{H}{\rm S}}/{\rm tr}(e^{-\beta\tdow{H}{\rm S}})$ assuming $Q$ and $\tdow{H}{\rm S}$ have no common eigenspace. To unravel the thermodynamical nature of the irreversible contribution of the LDME we can rewrite (\ref{LDME}) as 
\begin{equation}
  \dot{\rho} = -i\Qcommu{\tdow{H}{\rm S}}{\rho} + 
  \frac{1}{2}\sum_{\omega} h(\omega) \int_{0}^{1} d\lambda \, e^{-\lambda\beta\omega}[A^\dagger_\omega, \rho^\lambda[A_{\omega}, S(\rho)-\beta\tdow{H}{\rm S}]\rho^{1-\lambda}],
  \label{irrev_contrib_LDME_nonlin_form}
\end{equation}
showing that the von Neumann entropy operator $S(\rho)\!=\!-\ln{\rho}$ drives the dissipative dynamics. This formulation is equivalent to (\ref{LDME}) as can be shown using the KMS condition and the identity $\int_{0}^{1} d\lambda \, e^{-\lambda\beta\omega} \rho^\lambda[A_{\omega}, S(\rho)-\beta\tdow{H}{\rm S}]\rho^{1-\lambda} \!=\! A_\omega\rho - e^{-\beta\omega}\rho A_\omega$~\cite{FOTO2014_Lin_Rep,Taj2015_MDS}. 

Below, all relevant quantities are expressed in terms of the absorption rate $a(\omega)\! = \mbox{\small $h(-\omega)$}$ for $\omega\!>\!0$. For the sake of simplicity, we use for the qubit a parametrization in terms of the Pauli matrices~\footnote{The Pauli matrices $\sigma_x$, $\sigma_y$ and $\sigma_z$ together with the identity $\id$ form a complete basis for the space of $2\times2$ Hermitian matrices.} $\sigma_x$, $\sigma_y$ and $\sigma_z$ yielding $\tdow{H}{\rm S}\!=\!(\Delta/2)\sigma_z$ with $\Delta\!=\!E_2-E_1$ the energy gap while $Q \!=\! (1/2)(e^{i\theta}\sigma_{+} + e^{-i\theta}\sigma_{-})$, for $\sigma_{\pm}\!=\!\sigma_x \pm i\sigma_y$ and $\theta\in[0,2\pi)$, is dimensionless so that the units of energy are fully assigned to $\mbox{\small $h(\omega)$}$.

\section{Nonlinear extension} 
To go beyond the standard linear master equation (\ref{LDME}), exclusively producing exponential decays, and maintain the clear thermodynamical picture offered by (\ref{irrev_contrib_LDME_nonlin_form}) we make use of the nonlinear thermodynamic master equation~(NTME)~\cite{Taj2015_MDS}
\begin{equation}
  \begin{split}
  \dot{\rho}
   & = -i\Qcommu{\tdow{H}{\rm S}}{\rho} + \frac{1}{2}\sum_{\omega,\bar{\omega}}\sqrt{h(\omega)h(\bar{\omega})} 
   \int_{0}^{1} d\lambda \, e^{-\lambda\beta\frac{\omega+\bar{\omega}}{2}}[A^\dagger_\omega, \rho^\lambda[A_{\bar{\omega}}, S(\rho)-\beta\tdow{H}{\rm S}]\rho^{1-\lambda}].
   \end{split}
  \label{NTME}
\end{equation}
This truly nonlinear equation, inspired by a derivation~\cite{Grabert1982} as well as by thermodynamical~\cite{Ottinger2010_TLS_DHO,Ottinger2011} and statistical~\cite{Taj2010} arguments, generates a modular dynamical semigroup ensuring the preservation of the hermiticity, the trace and the positivity of $\rho$ as expected from a physical master equation~\footnote{In \cite{Taj2015_MDS} it is shown that the environment can be modelled as a quantum or classical object; here, we use a quantum heat bath.}. Moreover, it converges to the Gibbs state and gives rise to a positive entropy production. Beside, the NTME~(\ref{NTME}) gives back the LDME~(\ref{LDME},\ref{irrev_contrib_LDME_nonlin_form}) in the WCL, asymptotically describing the exact Hamiltonian dynamics in the long time limit~\cite{Taj2008}, by applying the time-averaging procedure mentioned in \cite{Taj2010}. The NTME~(\ref{NTME}) carries a sum over two sets of Bohr frequencies as master equations in absence of the secular approximation such as the Bloch-Redfield master equation~\cite{Breuer2002}. Such a contribution notably produces fast-oscillating terms but here without spoiling the positivity of the density matrix.

Pechukas~\cite{Pechukas1995} and Romero~\cite{Romero2004} noted that beyond the WCL nothing forbids to have an equation for the reduced system which is nonlinear with respect to the state. Indeed, the nonlinearity of a reduced system typically arises by  eliminating the irrelevant degrees of freedom of the total system's density matrix using generalized Nakajima-Zwanzig methods~\cite{Willis1974,Grabert1974,Grabert1976,Grabert1977,Linden1998}. Of course, the full system evolves under a linear von Neumann equation. Alternatively, it is  naturally produced from maps constructed over a physical domain of states~\cite{Dominy2015}, for example, due to the system's preparation~\cite{Stelmachovic2001}. Note also that nonlinear Markovian semigroups for open quantum system extending the Lindblad theorem have been worked out by Alicki~\cite{Alicki1983} and Belavkin~\cite{Belavkin1989}, e.g.~to study quantum Boltzmann or Hartree-type of equations~(for further explorations see \cite{Alicki2009,Kolokoltsov2010,Lugiewicz2013}). The necessity to go beyond complete positivity was pointed out~\cite{Majewski1990,Alicki1990} mainly because the standard definition of complete positivity is only physically relevant when associated to a linear dynamics~\cite{Czachor1998}. In this respect, the present nonlinear Markovian reduced dynamics is pertinent from a physical perspective.

\section{Near equilibrium} 
Given the nonlinear nature of the NTME, the various lifetimes can only be accessed through the first order susceptibility. As shown in \cite{FOTO2014_Lin_Rep} the latter is obtained by means of the fluctuation-dissipation theorem~(FDT) in the time-domain through the expression $\tdow{\chi}{AB}(t)\!=\!-\beta\,\partial_t{\rm tr}(A e^{\mathcal{L}t} \mathnormal{K}_\pi B)$ relative to the self-adjoint observables $A$ and $B$ as well as based on the linearization $\dot{\rho}\!=\!\mathcal{L}\rho$ of (\ref{NTME}) near equilibrium obtained by replacing
\begin{equation}
  \rho^\lambda[A_{\bar{\omega}},S(\rho)-\beta\tdow{H}{\rm S}]\rho^{1-\lambda} \to \pi^\lambda[A_{\bar{\omega}}, \mathcal{K}_\pi^{-1}\rho]\pi^{1-\lambda}.
  \label{NTME_linearization}
\end{equation}
Above, $\mathcal{K}_\pi^{-1}\rho \!=\! \int_0^\infty\!ds(\pi+s)^{-1}\rho\,(\pi+s)^{-1}$ is the inverse of $\mathcal{K}_{\pi} A \!=\! \int^1_0\!d\lambda\,{\pi}^\lambda A {\pi}^{1-\lambda}$ known as the equilibrium Kubo-Mori superoperator~\cite{Petz1994}. Note that, contrary to the thermodynamically robust NTME, its linearized version does not preserve positivity far away from equilibrium and one should limit its use to the linear response regime where the lifetimes/decay rates are defined~\footnote{Decay times have an experimental significance for long times near equilibrium~\cite{Sevian1989}. The decay rates obtained by fitting the polarizations far away from equilibrium are not equal to the one defined in the linear response regime; this is even more true with the NTME which is nonlinear~(\textit{i.e.}~distinct initial conditions can yield different decay rates; match is only recovered close to equilibrium).}.

To perform the calculations, it is advantageous to switch to the Liouville space~(see, \textit{e.g.}~chap.~3 of~\cite{Mukamel1995}) highlighted hereafter with a bold notation. Choosing the vector representation $\bm{\rho}\!=\!(\rho_{11},\rho_{12},\rho_{21},\rho_{22})^{T}$ for the density matrix coefficients in the Hamiltonian basis, the linearized NTME (\ref{NTME})--(\ref{NTME_linearization}) reads $\bm{\dot{\rho}}\!=\!\bm{\mathcal{L}}\bm{\rho}$. Defining the dimensionless temperature~$\tilde{\beta}\!=\!\beta\Delta$, we compute the $4\!\times\!4$~matrix~$\bm{\mathcal{L}}$ for zero/large pure dephasing associated to the decay rate $\Gamma_2^{*}\!=\! \frac{1}{2} a(0) {|\tdow{Q}{{11}}-\tdow{Q}{{22}}|}^2$. This implies that the populations $\rho_{11}$ and $\rho_{22}$ decouple from the coherence $\rho_{12}$, leading to the Liouville generator~\footnote{For large pure dephasing, $|\tdow{Q}{{11}}-\tdow{Q}{{22}}|$ is negligible in regard of ${|\tdow{Q}{{11}}-\tdow{Q}{{22}}|}^2$ producing the zero entries in $\bm{\mathcal{L}}$.}
\begingroup
\renewcommand*{\arraystretch}{1.2}
\begin{equation}
  \bm{\mathcal{L}} =
  \begin{pmatrix} 
    -x  & 0             	& 0             	& x\,\tupp{e}{\tilde{\beta}} \\
    0 	& -y+i\Delta   		& z\,e^{2i\theta} 	& 0 \\		
    0 	& z\,e^{-2i\theta} 	& -y-i\Delta  		& 0 \\
    x 	& 0             	& 0             	& -x\,\tupp{e}{\tilde{\beta}} \end{pmatrix} 
  \label{Liouville_matrix_simplified}
\end{equation}
\endgroup
depending on three real positive variables
\begin{eqnarray}
\label{real_param_NTME}
  x &=&  a(\Delta), \nonumber \\
  y &=& (1+\tupp{e}{\tilde{\beta}}) \, a(\Delta)/2 - \Gamma_2^{*}, \\
  z &=& (\tilde{\beta}/2) \, \tupp{e}{\tilde{\beta}/2} \, {\rm coth}(\tilde{\beta}/2) \, a(\Delta). \nonumber
\end{eqnarray}

\textit{Spectral analysis } The four eigenvalues of the generator (\ref{Liouville_matrix_simplified})--(\ref{real_param_NTME}) are simply $0$, $-\tdow{\Gamma}{1}$ and $-\tdow{\Lambda}{\pm}$ given by $\tdow{\Gamma}{1}\!=\! (1+\tupp{e}{\tilde{\beta}})\,x$ and $\tdow{\Lambda}{\pm}\!=\! y \pm \Omega$ with $\Omega\!=\!\sqrt{z^2-\Delta^2}$ being real or imaginary. From $\bm{\mathcal{L}}$ and its eigenvalues it naturally follows that $\tdow{\Gamma}{1}\!=\!1/\tdow{T}{1}$ is the energy relaxation rate. On the other hand, the \textit{real part} of $\tdow{\Lambda}{\pm}$ is associated to either one decoherence rate $\tdow{\Gamma}{2}\!=\!1/\tdow{T}{2}\!=\!y$, for $z<\Delta$, or two decoherence rates $\tdow{\Gamma}{2\pm}\!=\!1/\tdow{T}{2\pm}\!=\!\tdow{\Lambda}{\pm}$ for $z>\Delta$. It is noteworthy that all decay rates are relative either to energy relaxation or decoherence, \textit{i.e.}~no $\tdow{\Gamma}{3}$ coupling populations and coherence appears in the model. The transition from one to two decay rates happens for $z>\Delta$, or equivalently, past the absorption rate $a(\Delta)$ threshold
\begin{equation}
  a_{\rm thr} = \Delta\,\tupp{e}{-\tilde{\beta}/2}\,{\tanh{(\tilde{\beta}/2)}}/{(\tilde{\beta}/2)}
  \label{threshold_NTME_abs}
\end{equation}
obtained from (\ref{real_param_NTME}) with $z\!=\!\Delta$. Before the threshold, we observe a strong analogy with the LDME equally predicting a single decoherence time~\footnote{Setting the parameter $z=0$ in the linearized NTME generator (\ref{Liouville_matrix_simplified})--(\ref{real_param_NTME}) yields the famous LDME generator.}. The ``branching'' of the decoherence times beyond the threshold is always reached for low enough temperatures~($a_{\rm thr}\!\to\!0$ for $\tilde{\beta}\!\gg\!{1}$) while for very high temperatures~($a_{\rm thr}\!\to\!\Delta$ for $\tilde{\beta}\!\ll\!{1}$) it requires $a(\Delta)\!\sim\!\Delta$ a strong coupling regime~\footnote{To model the strong coupling regime one could, for example, include short-time effects in the spectral function.}. It is remarkable that such a temperature-dependent transition from one to two decoherence times, which are associated to a biexponential decay of the coherence, was reported in low temperature experiments of nuclear spins' impurities in silicon crystals~\cite{Jarvinen2014,Gumann2014} or InGaAs quantum dots~\cite{Borri2003}.

\section{Susceptibility analysis} 
We note that the bifurcation phenomenon has simple and striking consequences on the susceptibility $\tdow{\chi}{DD}(t)$ associated to the dipole operator $D \!=\! \mu(e^{i\psi}\sigma_{+} + e^{-i\psi}\sigma_{-})$ with $\mu\!>\!0$. Switching to the frequency domain the aforementioned FDT becomes $\tdow{\chi}{AB}(\nu) \!=\! -\beta\, \bm{A} \!\cdot\! {\bm{\mathcal{L}}} \tupp{(\bm{\mathcal{L}} - i\nu\bm{1})}{-1} \, \bm{\mathcal{K}_\pi} \, \bm{B}$, where the scalar product is defined as $\bm{X}\!\cdot\!\bm{Y}=\sum_{1}^{4}x_{i}^{*}y_i$. $\bm{A}$ and $\bm{B}$ are again two self-adjoint observables now expressed as $4\!\times\!1$ vectors whereas $\bm{\mathcal{K}_\pi}$ is a $4\!\times\!4$ diagonal Kubo matrix evaluated at equilibrium whose non-zero elements read $\{\,\tupp{(1+\tupp{e}{-\tilde{\beta}})}{-1} ; \tupp{\tilde{\beta}}{-1}\tanh{(\tilde{\beta}/2)} ;\tupp{\tilde{\beta}}{-1}\tanh{(\tilde{\beta}/2)} ; \tupp{(1+\tupp{e}{\tilde{\beta}})}{-1}\,\}$ and $\bm{1}$ is a $4\times4$ identity matrix. The dipolar susceptibility arising from the FDT then reads
\begin{equation}
  \tdow{\chi}{DD}(\nu) = (\mu^2/\Delta)\tanh{(\tilde{\beta}/2)} 
  \sum_{\pm} \left( 1 \pm \frac{z}{\Omega} \cos{(2(\theta-\psi))} \right) \frac{\tdow{\Lambda}{\pm}}{\tdow{\Lambda}{\pm}+i\nu}.
  \label{susceptibility_DD_NTME}
\end{equation}
Prior to the bifurcation, the absorption ${\rm Im}(\tdow{\chi}{DD}(\nu))$ can \textit{essentially} be approximated by a superposition of two Lorentzians, centered at ${\rm Im}(\tdow{\Lambda}{\pm})$, with a unique linewidth equal to $\tdow{\Gamma}{2}$ for frequencies $\nu\!\simeq\!{\rm Im}(\tdow{\Lambda}{\pm})$, \textit{i.e.}~close to the resonances. After the bifurcation, however, the two resonances are both located at the frequency $\nu\!=\!{0}$, each of which is characterized by its own linewidth $\tdow{\Gamma}{2\pm}$, being neither of Lorentzian nor Gaussian type. Indeed, we have a superposition of two non-Lorentzian functions ${\rm Im}(\tdow{\chi}{DD}(\nu))\!=\!\sum_{\pm} c_{\pm}\nu/(\tdow{\Lambda}{\pm}^2\pm\nu^2)$ with suitable real coefficients $c_{\pm}$~\footnote{The approximation $\nu\!=\!0$ at the numerator cannot be used to restore a combination of Lorentzian in case $c_{\pm}$ and $\tdow{\Lambda}{\pm}\!=\!0$ so that the profile is truly non-Lorentzian.}. Notably, non-Lorentzian lineshapes were measured for nuclear spins~\cite{Sasaki2005}, a wide range of quantum dots~\cite{Birkedal2001,Borri2001,Borri2003,Vagov2004,Borri2007}, NV centers~\cite{Balasubramanian2009,Smith2011} and for a quantum well~\cite{Stich2007}. Moreover, if one of the two contributions is much narrower than the other, a peak which closely resembles a zero-phonon line arises~\cite{Grosse2008,Deveaud2009,Accanto2012}. Our analysis shows that the NTME offers straightforward modeling tools for these phenomena, impossible to describe with the corresponding LDME. It also indicates that non-Markovian effects~\cite{Chang1993} are not necessarily required to produce non-Lorentzian profiles~\footnote{The NTME has been designed to be Markovian and we checked this fact with a non-Markovianity measure~\cite{Breuer2009}.}. 

\section{Dynamical analysis} 
\begin{figure}
  \includegraphics[scale=1.5]{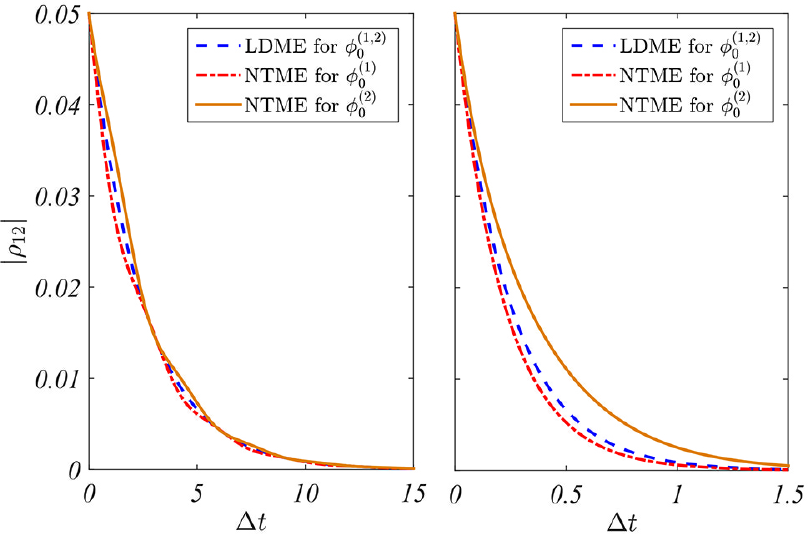}
  \caption{Time-evolution \textit{before}~(left~panel,~$a(\Delta)\!=\!0.002\Delta$) and \textit{after}~(right~panel,~$a(\Delta)\!=\!0.02\Delta$) the threshold $a_{\rm thr}=0.016\Delta$ at large pure dephasing~(${|\tdow{Q}{{11}}-\tdow{Q}{{22}}|}^2=1.3$) and low temperature~($\tilde{\beta}=6$). We compare initial states with distinct orientations $\tsubb{\phi}{0}{(1)}\!=\!\frac{\pi}{5}$ and $\tsubb{\phi}{0}{(2)}\!=\!\tsubb{\phi}{0}{(1)}+\frac{\pi}{2}$ while $\tsubb{r}{0}{(1,2)}\!=\!0.05$.}
  \label{fast_slow_Bloch_fig}
\end{figure}

To fully capture the implications of the two decoherence times, we solve the system of equations generated by the Liouville generator (\ref{Liouville_matrix_simplified}) for ${\rho}_{12}$ with the initial condition ${\rho}_{12}(0) \!=\! \tdow{r}{0}\,e^{i\tdow{\phi}{0}}$, whose modulus must respect ${r_0}^2\leq{\rho}_{11}(0)\,{\rho}_{22}(0)$, yielding
\begin{equation}
  \begin{split}
     {\rho}_{12}(t) & = \frac{\tdow{r}{0}}{2\Omega}
			\left\{ e^{-t\tdow{\Lambda}{+}} \left[ (i\Delta+\Omega)e^{i\tdow{\phi}{0}} + z e^{i(2\theta-\tdow{\phi}{0})} \right] \right\} \\
		    & - \frac{\tdow{r}{0}}{2\Omega}
			\left\{ e^{-t\tdow{\Lambda}{-}} \left[ (i\Delta-\Omega)e^{i\tdow{\phi}{0}} + z e^{i(2\theta-\tdow{\phi}{0})} \right] \right\}.
  \end{split}
  \label{sol_sub_lin_system_rho_offdiag}
\end{equation}
Increasing $a(\Delta)$ beyond the threshold (\ref{threshold_NTME_abs}) induces a transition from a single oscillating exponential decay~(\textit{i.e.}~$\tdow{\Lambda}{\pm}$ is complex) towards a non-oscillating biexponential one~(\textit{i.e.}~$\tdow{\Lambda}{\pm}$ is real) associated to a short $\tdow{T}{2+}$ and a long $\tdow{T}{2-}$ decoherence time. Both cases are presented in fig.~\ref{fast_slow_Bloch_fig}. The dependence on the orientation $\phi_0$ of the initial Bloch vector, displayed by the NTME but not by the LDME, has been qualitatively measured in presence of a spin bath~\cite{Stanwix2010,Shin2013}. Moreover, the crossover from an oscillatory towards a biexponential damping has been observed for the spin dynamics in a 2D electron gas~\cite{Weber2007}. More generally, a biexponential decay was measured in spins~\cite{Sasaki2005,Shankar2010,Jarvinen2014,Gumann2014}, NV centers~\cite{Beha2012,Grezes2014}, light-harvesting complexes~\cite{Muller2010,SchlauCohen2012} or in various quantum dots~\cite{Birkedal2001,Borri2001,Borri2003,Vagov2004,Borri2007,Robert2012,Birindelli2014}~and~\cite{Accanto2012,Raino2012}. To understand which physical mechanisms produce these unorthodox decays, one should investigate each setup thoroughly. This is done, for example, for GaAs quantum dots in chap.~4 of \cite{Deveaud2009} where the origins of the non-exponential decay and non-Lorentzian profiles are discussed~(spin-orbit coupling or electron-phonon/hyperfine interactions and so on)~\footnote{Note that the full decay pattern is much richer than what the NTME predicts in linear response regime, \textit{e.g.}~see fig.~14 on p.~93 of \cite{Deveaud2009} for a detailed overview of short, intermediate and long times.}. With these insights, one could get a reasonable model by specifying the parameters of the NTME.

In general, one would like to obtain a time evolution solely driven by the longest decay time $\tdow{T}{2-}$ to preserve the coherence as long as possible. To this end, we set the first term in eq.~(\ref{sol_sub_lin_system_rho_offdiag}) equal to zero yielding
\begin{equation}
  -(i\Delta+\Omega)\,z^{-1} = e^{2i(\theta-\phi_0)}.
  \label{condition_enhanced_longest_T2}
\end{equation}
To satisfy this condition for a given setup, one should start from an initial state with $\phi_0$ equal to the critical Bloch angle $\phi_c \!= \! \theta-{\rm arccos}(-\Omega/z)/2$. Thus, it is possible to generate a prolonged time-evolution of the coherence by taking an initial Bloch vector being oriented along the critical angle $\phi_c$, \textit{i.e.}~producing an \textit{optimal initial state}. This could drastically increase the coherence time being the cardinal resource to perform  quantum gate operations. Note that there exists many strategies to prepare such an initial state~\cite{Verstraete2009,Fleming2011} and that an iterative algorithm to track optimal Bloch vectors has been developed for materials doped with rare-earth ions~\cite{Heinze2014}.

\section{$T_1T_2$ ratio} 
\begin{figure}
  \includegraphics[scale=1.5]{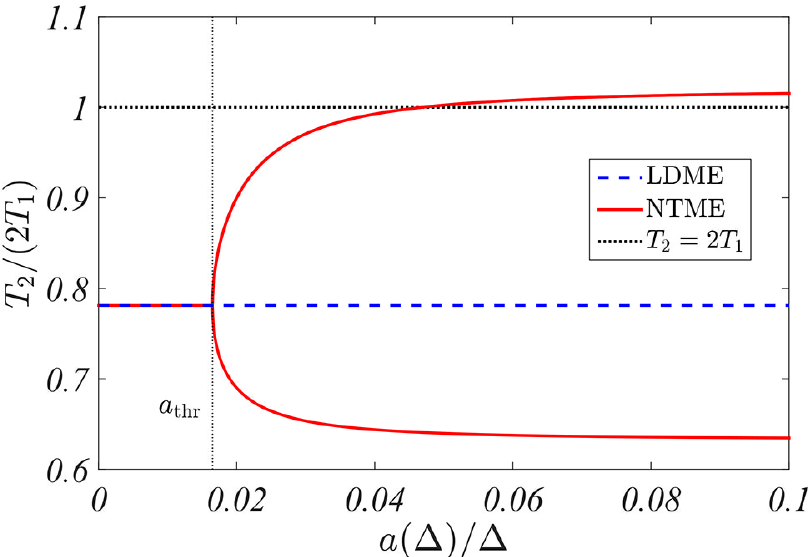}
  \caption{Bifurcation for the absorption $a(\Delta)>a_{\rm thr}$. We have used the same set of parameters as for fig.~1.}
  \label{ratio_T1T2_fig}
\end{figure}

Beyond the threshold (\ref{threshold_NTME_abs}) the linearized NTME predicts that the decoherence time $\tdow{T}{2-}$ is no longer restricted by the energy relaxation time $T_1$, even in the weak coupling regime, as illustrated in fig.~\ref{ratio_T1T2_fig}.
The $T_2 \leq 2T_1$ bound violation was already obtained by Laird and co-workers~\cite{Laird1991a,Laird1991b,Laird1994} but without a bifurcation and with positivity issues~\cite{Sevian1989,Andreozzi1992,Reichman1996}. On the contrary, our approach provides a thermodynamically and statistically safe way to overcome the bound. Furthermore, a merely formal nonlinear master equation obtained in \cite{Wonderen2000}, being positive but not completely positive, also does not meet the inequality. Indeed, there exists a one-by-one link between this inequality and complete positivity~\cite{Kimura2002}. As mentioned by Pechukas~\cite{Pechukas1994} as well as by Shaji and Sudarshan~\cite{Shaji2005}, complete positivity needs not to be a physical requirement in spite of its mathematical attractiveness.

\newpage 

\section{Conclusion and perspective} 
The LDME is over-restricted by complete positivity and does not allow to interpret numerous experimental results. To model realistic physical setups, one should give up linearity and accept nonlinearity as a feature of the reduced dynamics outside of the WCL~\cite{Pechukas1995}. The NTME gives rise to a bifurcation beyond a threshold, associated with a biexponential decay, allowing to drastically prolong the coherence for optimal orientation of the initial Bloch vectors. Moreover, according to our analysis nothing forbids to overcome the bound $T_2 \leq 2T_1$ although its experimental realization would clearly be challenging. Such conclusions could be made directly from the Liouville matrix (\ref{Liouville_matrix_simplified}), and its associated Bloch equation, \textit{per se} without any reference to the NTME (\ref{real_param_NTME}). Moreover, also other master equations can be written in this form~\footnote{Look, for example, at Laird's eq.~(A1)-(A4) in the Appendix of \cite{Laird1991a} where they neglected the case of $\Omega\!=\!\sqrt{z^2-\Delta^2}$ being real.}. For example, we observe similar results with one of the equations proposed in \cite{Taj2015_MDS} obtained by introducing a dynamical time coarse-graining.

Remarkably, the observed ultralong coherence is not limited to a single qubit and could be ``scaled-up'' since biexponential decays were measured in epitaxial quantum dot arrays~\cite{Ardavan2003}. Thus, it could allow to avoid entanglement sudden-death as well as to enhance revival\cite{Xu2010a,Sun2012,Aolita2015}. Away from the linear response regime, the decay pattern becomes richer and one then finds initial optimal states either by screening strategies or applying search/reinforcement learning algorithms~\cite{Sun2014}. Moreover, at intermediate pure dephasing, where the populations and coherences are interwoven, the NTME predicts population beating and coherence revival which have been observed in photosynthetic complexes~\cite{Lee2007,Scholes2011} or quantum kicked rotors~\cite{Pattanayak1997,Wu2009}. Taking all this into account, we hope that the present findings will provide some leads to develop ground-breaking nanoscale devices in the near future.

\begin{acknowledgments}
We would like to thanks R.~Alicki and G.~M.~Graf for stimulating discussion concerning the nonlinear nature of the thermodynamic master equation.
\end{acknowledgments}

\newpage


%

\end{document}